\documentclass[twocolumn, showpacs, showkeys,letter]{revtex4}
\usepackage{graphicx}
\usepackage{amsmath, amsthm, amsfonts,amssymb}

\begin{document}

\title{Topological Properties of the Minimal Spanning Tree in the Korean and American Stock Markets}

\author{Cheoljun Eom}
\affiliation{Division of Business Administration, Pusan National University, Busan 609-735, Korea}
\author{Gabjin Oh}
\email{gq478051@postech.ac.kr}
\affiliation{NCSL, Department of Physics, Pohang University of Science and Technology, Pohang, Gyeongbuk, 790-784, Korea}
\author{Seunghwan Kim}
\affiliation{NCSL, Department of Physics, Pohang University of
Science and Technology, Pohang, Gyeongbuk, 790-784, Korea}
\affiliation{Asia Pacific Center for Theoretical Physics, Pohang,
Gyeongbuk, 790-784, Korea}

\begin{abstract}
We investigate a factor that can affect the number of
links of a specific stock in a network between stocks created by the
minimal spanning tree (MST) method, by using individual stock data
listed on the S\&P500 and KOSPI. Among the common factors mentioned
in the arbitrage pricing model (APM), widely acknowledged in the
financial field, a representative market index is established as a
possible factor. We found that the correlation distribution,
$\rho_{ij}$, of 400 stocks taken from the S\&P500 index shows a very
similar with that of the Korean stock market and those deviate from
the correlation distribution of time series removed a nonlinearity
by the surrogate method. We also shows that the degree distribution
of the MSTs for both stock markets follows a power-law distribution
with the exponent $\zeta \sim$ 2.1, while the degree distribution of
the time series eliminated a nonlinearity follows an exponential
distribution with the exponent, $\delta \sim 0.77$. Furthermore the
correlation, $\rho_{iM}$, between the degree k of individual stock,
$i$, and the market index, $M$, follows a power-law distribution,
$\langle \rho_{iM}(k) \rangle \sim k^{\gamma}$, with the exponent
$\gamma_{\textrm{S\&P500}} \approx 0.16$ and
$\gamma_{\textrm{KOSPI}} \approx 0.14$, respectively. Thus,
regardless of the markets, the indivisual stocks closely related to
the common factor in the market, the market index, are likely to be
located around the center of the network between stocks, while those
weakly related to the market index are likely to be placed in the
outside.
\end{abstract}

 \maketitle

\section{Introduction}
Recently, researchers from diverse disciplines are showing great
interest in the topological properties of networks. In particular,
the network observed in natural and social science shows a variety
of properties different from those of random graphs
\cite{watts99,barabasi02} . The economic world, known as having the
most complex structure among them, evolves through the
nonlinear-interaction of the diverse heterogeneous agents. The stock
market is a representative example. The stock prices of individual
companies are formed by a complex evolution process of diverse
information generated in the market and these have strong
correlations with each other by the common factors in the market
\cite{Farrell1974,Ross1976,Chen1986}. In other words, individual
stocks are connected with each other and companies with the same
properties tend to be grouped into a community. To investigate these
properties, Mantegna.{\em et al.} proposed the minimal spanning tree
(MST) method, which can be used to observe the grouping process of
individual stocks transacted in the market, on the basis of the
correlation of stocks \cite{Mantegna1999a}. Mantegna constructed the
stock network visually using the MST method and found that this
generated network between stocks has an economically significant
grouping process \cite{Mantegna1999a,Mantegna1999b}.

The studies of the past several years showed that the degree
distribution of the network created by the MST method follows a
power-law distribution with the exponent $\xi$ $\approx 2$
\cite{Kim2002,Vandewalle2000}. That is, most individual companies in
the stock market have a small number of links with other stocks,
while a few stocks have a great number of connections. However the
KOSPI200 companies of the Korean stock market, one of the emerging
market, does not follow a power-law distribution and for the
American stock market, S\&P500, the relation between market
capitalization and $|q|$, the influence strength (IS), has a
positive correlation, while the KOSPI200 has no correlation
\cite{Jung2006}. Moreover, previous results showed that the network
of individual stocks tends to gather around the companies of a
homogeneous industrial group
\cite{Bonanno2000a,Bonanno2000b,Bonanno2003,Onnela2003}.
Furthermore, the stocks forming a Markowitz efficient portfolio in
the financial field are almost located at the outside of the network
\cite{Onnela2003,Onnela2003b}. However, until recently, studies on
the possible factors important in determining the number of linkages
with other stocks in the network between stocks were insufficient.

In order to investigate the topological properties in the network of
the stock market, we used the MST method introduced by Mantegna{\em
et al.} We also consider the market index in terms of a possible
factor that can affect the number of links of a specific stock with
other stocks in the network created by the MST method. We used the
data of 400 individual companies listed on the S\&P500 index from
January 1993 to May 2005, and 468 individual companies listed on the
KOSPI from January 1991 to May 2003.

We found that the correlation distribution, $\rho_{ij}$, of the 400
stocks in S\&P500 index shows a very similar with that of the KOSPI
and those deviate from the correlation distribution of time series
removed a nonlinearity by the surrogate method introduced by J.
Theiler {\em et al.} \cite{Theiler1992} for both stock markets. We
also found that the degree distribution of the network, like those
from previous research, follows a power-law distribution with the
exponent $\zeta_{\textrm{S\&P500, KOSPI}} \approx 2.1$ for both the
Korean and American stock markets. In order to observe the possible
factor in determing the degree k on the MST network, we calculate
the cross-correlation, $\rho_{iM}$, between a individual stock and
market index for both stock markets. We also found that the
cross-correlation, $\rho_{iM}$ between the market index and the
companies with the degree k follows a power-law distribution,
$\langle \rho_{iM}(k) \rangle \sim k^{\gamma}$, where the exponents
are calculated to be $\gamma_{\textrm{S\&P500}} \sim$ 0.16,
$\gamma_{\textrm{KOSPI}} \sim$ 0.14. In other words, individual
stocks having many connections with other stocks in the network
obtained by the MST method are more highly related to the market
index than those having a comparatively small number of links.

In the next section, we describe the financial data used in this
paper. In Section 3, we introduce the methodology. In Section 4, we
present the results in this investigation. Finally, we end with the
summary.

\section{Data}
We used 400 individual daily stocks data from January 1993 to May
2005 taken from individual stocks listed on the S\&P500 index of the
American stock market (from the Yahoo website) and 468 individual
daily stocks data from January 1991 to May 2003 taken from
individual stocks listed on the KOSPI of the Korean stock market
(from the Korean Stock Exchange). In order to investigate the
possible factors determined the number of links of an individual
stock in the network, we used the $S\&P 500$ and KOSPI index with
the same period as individual stocks, respectively. We used the
normalized returns, $R_t$, by the standard deviation, $\sigma(r_t)$,
after calculating the returns from the stock price, $P_t$, by the
log-difference, $r_t \equiv \ln P_{t} - \ln P_{t-1}$, as in previous
studies and defined as follow

\begin{equation}
{R_t} \equiv  \frac{\ln{P_{t}}- \ln{P_{t-1}}}{\sigma(r_t)},\\
\label{e1}
\end{equation}
where $\sigma(r_t)$ is the standard deviation of the return.

\section{Methodology}
\label{sec:METHODOLOGY} We make the network by using stocks listed
on the S\&P500 and KOSPI, respectively, through the MST method
proposed by Mantegna {\em et al.} As the MST method makes the
network based on the correlation of stocks, the cross-correlation of
stocks listed on the S\&P 500 and KOSPI stock markets, respectively,
is calculated as follows

\begin{equation}
 \rho_{ij} = \frac{\langle R_{i}R_{j} \rangle - \langle R_{i}\rangle \langle R_{j}\rangle}{\sqrt{({\langle R_{i}^{2}\rangle}-{\langle R_{i}\rangle}^{2})
 ({\langle R_{j}^{2}\rangle}-{\langle R_{j}\rangle}^{2})    }}, \\
 \label{e2}
\end{equation}
where $\langle .\rangle$ means the mean value of the whole period
and the correlation lies within the range of $-1 \leq \rho_{ij} \leq
+1$. If $\rho_{ij}$ is 1, two time series have a complete
correlation and if $\rho_{ij}$ is -1, they have a complete
anti-correlation. In the case where $\rho_{ij}$ is 0, the
correlation of two time series is 0. On the basis of $\rho_{ij}$
calculated by Eq. \ref{e2}, the distance between nodes is calculated
as follows

\begin{equation}\label{e3}
d_{ij}= \sqrt{2(1-\rho_{i,j})}.
\end{equation}

In order to find out the correlation between the number of
connections of a individual stock with other stocks in the network
and the market index, we investigated the correlation, $\rho_{iM}$,
between the market index and individual stocks. Using the Eq.
\ref{e4}, we calculated the correlation, $\rho_{iM}$, between
returns of individual stocks, $R_i$, and the market index, $R_{M}$,
and defined as follow

\begin{equation}\label{e4}
 \rho_{iM} = \frac{\langle R_{i}R_{M} \rangle - \langle R_{i}\rangle \langle R_{M}\rangle}{\sqrt{({\langle R_{i}^{2}\rangle}-{\langle R_{i}\rangle}^{2})
 ({\langle R_{M}^{2}\rangle}-{\langle R_{M}\rangle}^{2})    }}. \\
\end{equation}

\section{Results}
\label{sec:RESULTS}

\begin{figure}

\includegraphics[height=8cm, width=8cm]{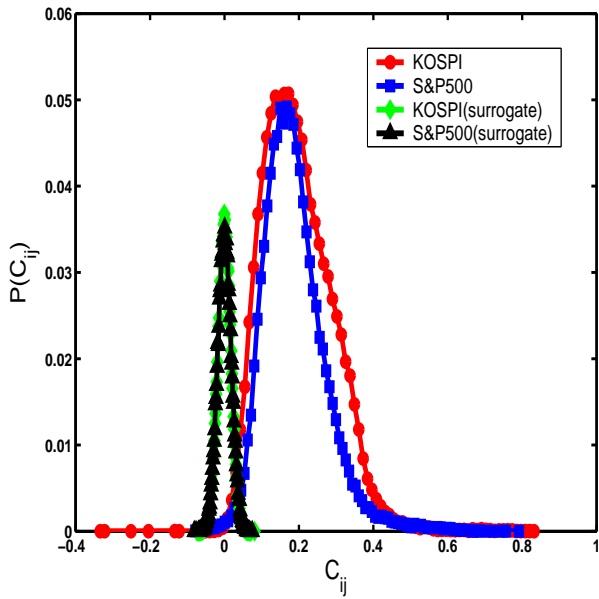}

\caption[0]{The PDF of the cross-correlation $\rho_{ij}$ between the
stocks listed on for the S\&P500 and KOSPI stock markets. The red
(circle), blue (square), green (diamond), and black (triangle)
denotes the KOSPI, S\&P500, KOSPI (surrogate), and S\&P500
(surrogate) stock market, respectively.} \label{fig1}
\end{figure}

\begin{figure}

\includegraphics[height=5cm, width=4cm]{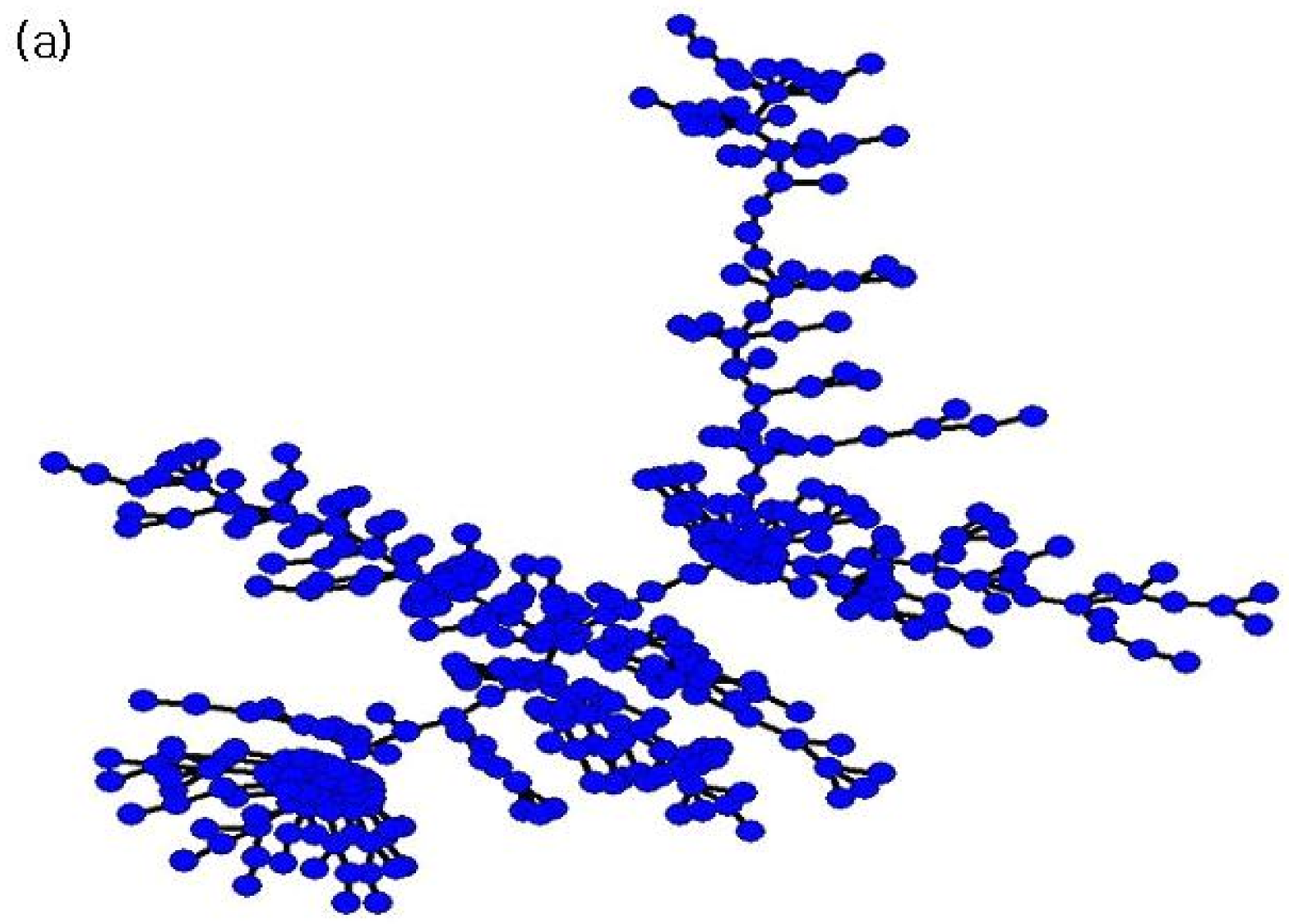}
\includegraphics[height=5cm, width=4cm]{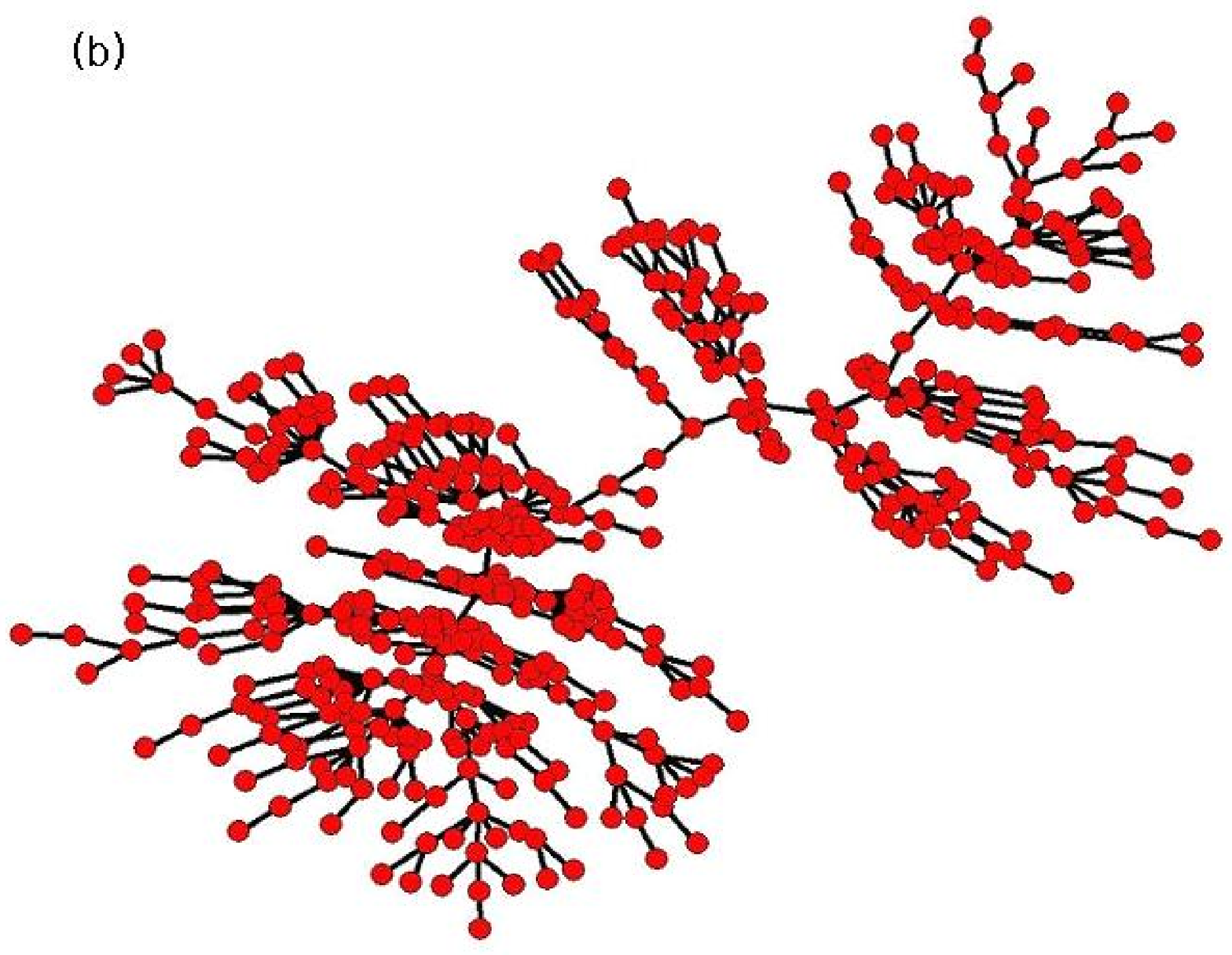}
\includegraphics[height=5cm, width=4cm]{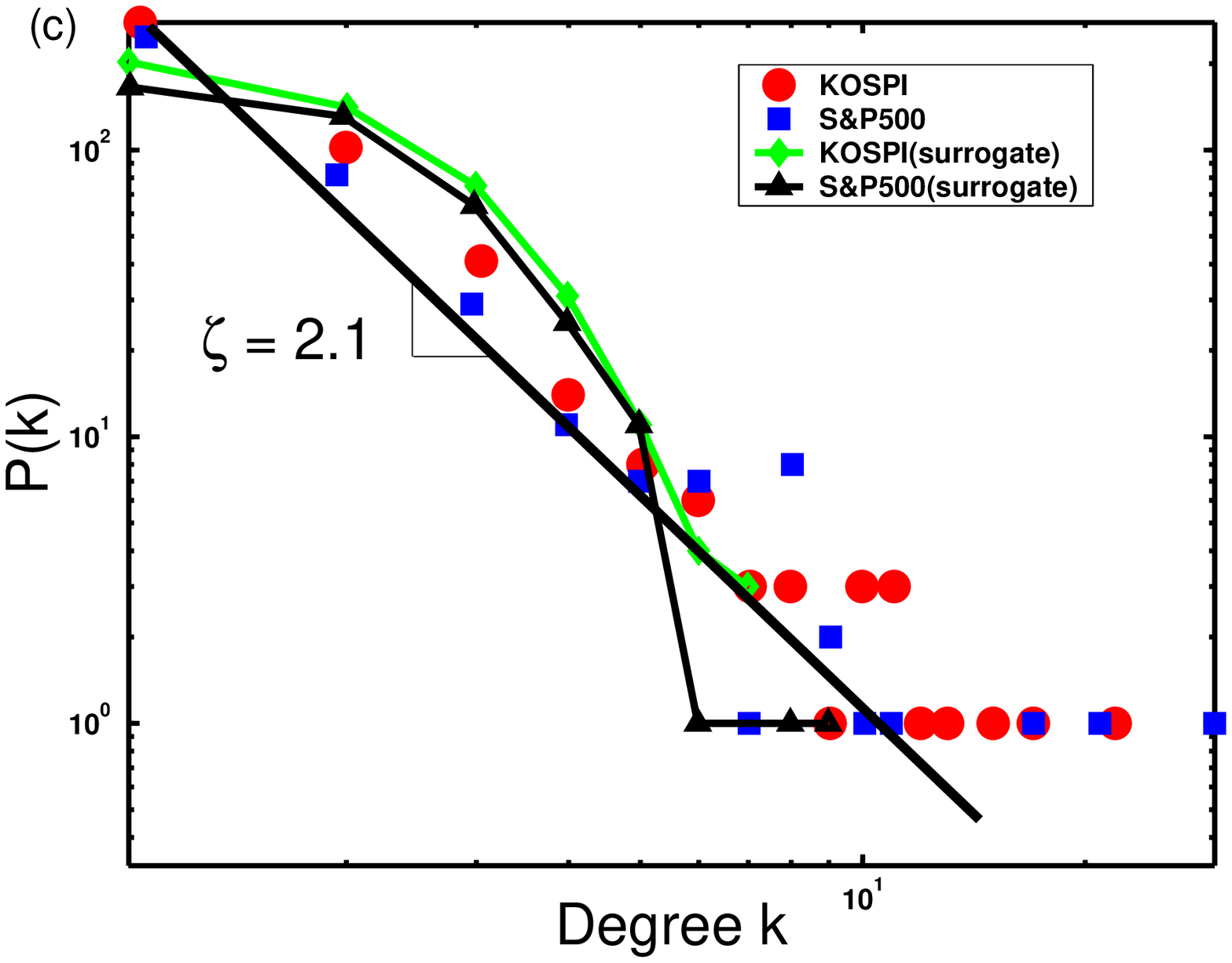}
\includegraphics[height=5cm, width=4cm]{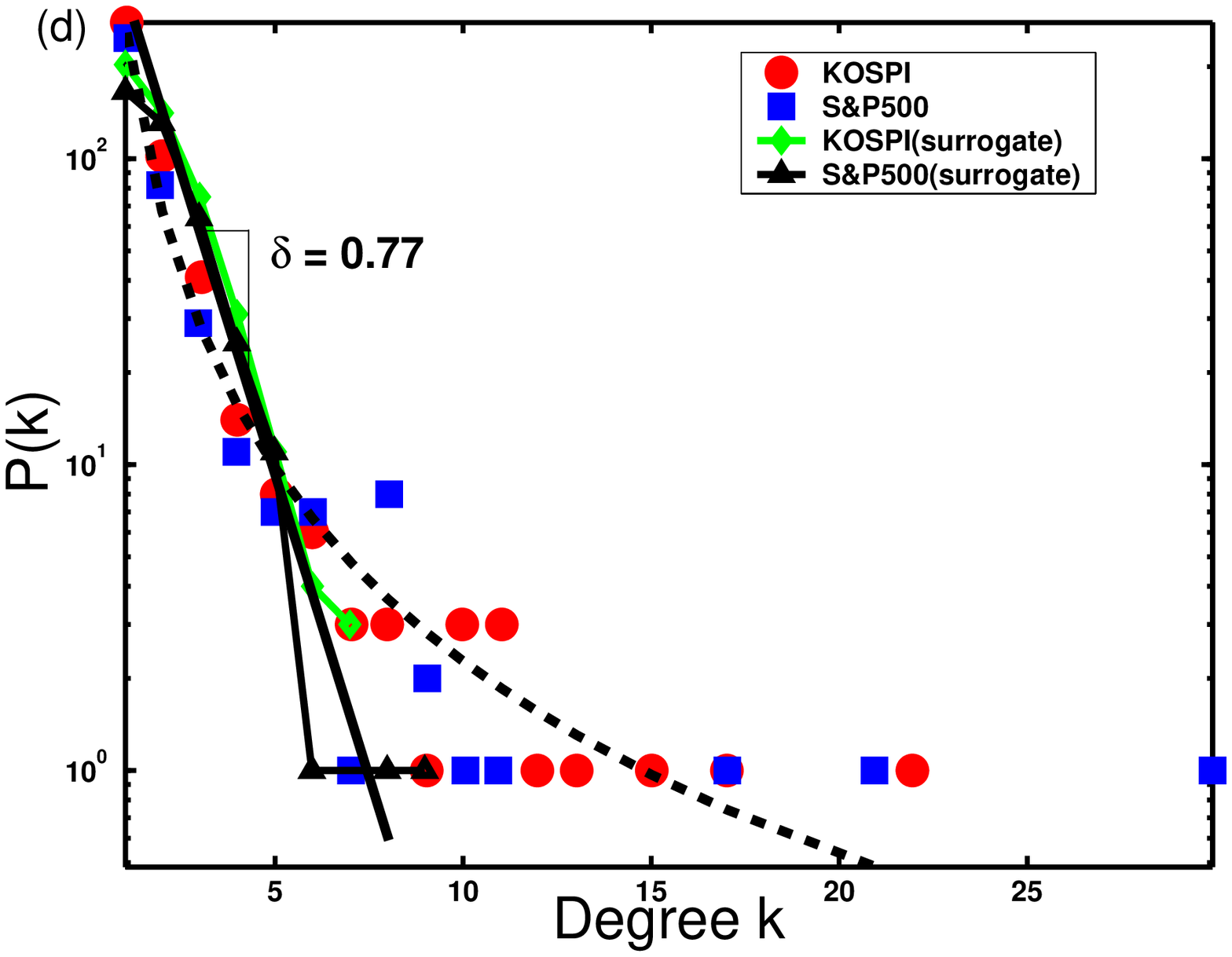}

\caption[0]{(a) MST structure composed of daily returns of 400
individual companies lised on the S\&P500 index from 1993 to 2005,
and (b) MST structure composed of the daily returns of 468
individual corporations listed on the KOSPI from 1991 to 2003. (c)
shows the degree distribution of the MST structure composed of
individual companies on the S\&P500 index and KOSPI. The degree of
both the S\&P500 index and KOSPI follows a power law distribution
with the exponent of $\zeta \sim 2.1$. The red (circle), blue
(square), green (diamond), and black (triangle) denotes the KOSPI,
S\&P500, KOSPI (surrogate), S\&P500 (surrogate), respectively.}
\label{fig2}
\end{figure}

In this section, using the MST method we investigated the network
properties of 400 individual stocks listed on the S\&P500 index from
January 1993 to May 2005, and 468 individual stocks listed on the
KOSPI from January 1991 to May 2003, respectively.

First, we analyze the distribution of the correlation matrix,
$\rho_{ij}$, for individual stocks in the S\&P500 index and KOSPI.
In Fig. \ref{fig1}, we shows the distribution of the correlation
matrix for both stock markets. The blue (square), red (circle),
green (diamond), and black (triangle) indicates the S\&P500, KOSPI,
S\&P500 (surrogate), KOSPI (surrogate) stock markets, respectively.
We find that the correlation distribution, $\rho_{ij}$ between
stocks in the S\&P500 index shows a very similar with that between
stocks in the KOSPI and those deviate from the correlation
distribution of time series created by the surrogate method. In
order to estimate the topology structure of both stock markets, we
employ the MST method proposed by the Mantegna {\em et al.} Using
the whole period data of the Korean and American stock markets, we
present the network structure calculated by the MST method and its
the degree distribution for both the stock markets. Fig. \ref{fig2}
shows the network structures between stocks generated by the MST
method, using individual stock data lised on the American and Korean
stock market, respectively and plot the degree distribution, $P(k)$,
of the MST networks for both stock markets. In Fig. \ref{fig2}, (a)
and (b) display the MST structure composed of daily returns of the
individual companies lised on the S\&P500 index and KOSPI,
respectively and (c) and (d) show the degree distribution of the MST
structure for both stock markets in the log-log and linear-log plot.
The red (circle), blue (square), green (diamond), and black
(triangle) indicates the KOSPI, S\&P500, KOSPI (surrogate), S\&P500
(surrogate), respectively and the notation (surrogate) denotes the
corresponding surrogate data.

We find that the degree distribution for both stock markets follows
a power law distribution with the exponent $\zeta \sim 2.1$, while
the degree distribution of MST network of the time series created by
the surrogate method \cite{Theiler1992} does follows an exponential
distribution with $\delta \sim 0.77$. Thus, as the results finding
in the complex network such as the internet, WWW, protein-protein
interaction, and so on, there is a scale-free network property.
Therefore, the hubs existence in the financial markets means that
there are a dominant company which gives many influence to the other
stocks. In order to observe a possible factor determination the
degree of an individual stock on the MST network, we calculated the
correlation, $\rho_{iM}$, between an individual stock with the
degree $k$ and the stock market index for both stock markets.

\begin{figure}
\includegraphics[height=10cm, width=8cm]{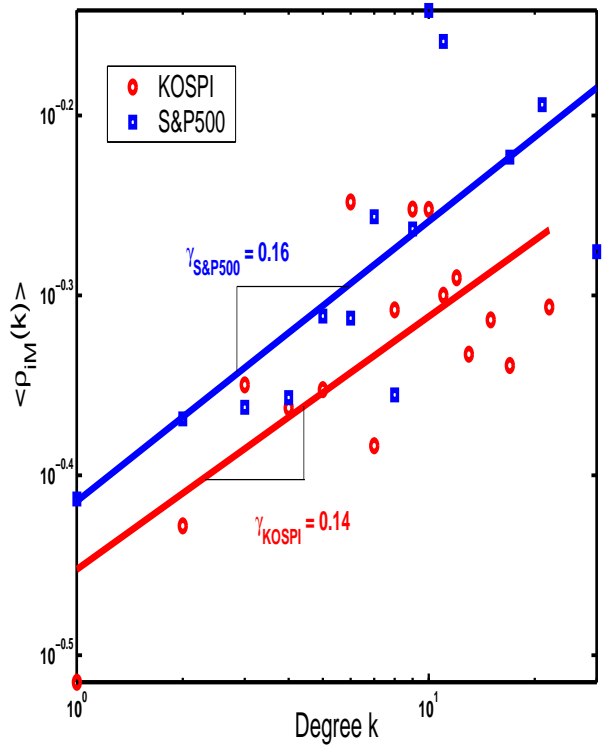}
\caption[0]{Distribution of the correlation, $\rho_{iM}$, between an
individual companies and market index for both the S\&P500 index and
KOSPI, respectively.} \label{fig3}
\end{figure}

In Fig. \ref{fig3}, we shows the distribution of the correlation,
$\rho_{iM}$, between stocks in the S\&P500 index and KOSPI,
respectively. The blue (square) and red (circle) indicates the
S\&P500 and KOSPI.

We find that the correlation, $\rho_{iM}$, between a stock with the
degree k, the number of connected with other stocks, and the market
index for both stock markets follows a power-law distribution,
$\langle \rho_{iM}(k) \rangle \sim k^{\gamma}$, with the exponent
$\gamma_{\textrm{S\&P500}} \approx 0.16$ and
$\gamma_{\textrm{KOSPI}} \approx 0.14$, respectively. Thus, through
these results, we found that the stocks having more links with other
stocks are more highly correlated with the market index than those
with relatively fewer connections. In other words, the stocks
closely related to the market index have a larger number of links
with other stocks and are likely to be located around the center of
the network of the stocks. On the other hand, the stocks poorly
related to the market index have fewer links with other stocks and
tend to be placed at the outside of the network of the stocks. Our
results suggest that the common factor such as the market index play
a important rules in terms of determing the networks in the
financial markets.

\section{Conclusions}
In this paper, in order to examine possible factors capable of
affecting the number of links that a specific stock has in relation
to other stocks in the network between stocks created using the MST
method, we carried out research using the market index, a
representative among multiple common factors mentioned in the
arbitrage pricing model (APM). We used 400 individual stocks listed
on the S\&P 500 index and 463 stocks listed on the KOSPI.

We found that the correlation distribution, $\rho_{ij}$, between
stocks in the S\&P500 index shows a very similar with that between
stocks lised on the KOSPI and those deviate from the correlation
distribution of time series removed a nonlinearity by the surrogate
method. We shows that the degree distribution in the network between
stocks obtained by the MST method for both stock markets follows a
power-law distribution with the exponent $\zeta_{\textrm{S\&P500,
KOSPI}} \sim$ 2.1, while the degree distribution from the time
series eliminated a nonlinearity follows an exponential distribution
with the exponent, $\delta_{\textrm{S\&P500 (surrogate), KOSPI
(surrogate)}} \sim 0.77$. In order to investigate a factor
determining the degree k on the MST network, we used the market
index for both stock markets. We found that in the degree
distribution, the correlation, $\rho_{iM}$, between the degree k,
the number of links, and the market index for both stock markets
follows a power-law distribution, $\langle \rho_{iM}(k) \rangle \sim
k^{\gamma}$, with the exponent $\gamma_{\textrm{S\&P500}} \approx
0.16$ and $\gamma_{\textrm{KOSPI}} \approx 0.14$, respectively. In
other words, the stocks having the intimate relation with the market
index have a larger number of links, while the stocks poorly related
to the market index have fewer links. According to above finding
results, we imply that the degree $k$ as most important quantity to
describe the network topology, has a closely relation with the
common factors such as market index.

\begin{acknowledgments}
This work was supported by the Korea Research
Foundation funded by the Korean Government (MOEHRD)
(KRF-2006-332-B00152), and by a grant from the MOST/KOSEF to the
National Core Research Center for Systems Bio-Dynamics
(R15-2004-033), and by the Korea Research Foundation
(KRF-2005-042-B00075), and by the Ministry of Science \& Technology
through the National Research Laboratory Project, and by the
Ministry of Education through the program BK 21.
\end{acknowledgments}

\end{document}